# Hydrological Cycle in the Danube basin in present-day and XXII century simulations by IPCCAR4 global climate models

Valerio Lucarini[*] (1,2), Robert Danihlik (2) Ida Kriegerova (2), Antonio Speranza (2,3)

1. Department of Physics, University of Bologna, Bologna, Italy
2. CINFAI, Camerino Italy
3. Department of Mathematics and Computer Sciences, University of Camerino, Camerino, Italy

## Abstract

We present an intercomparison and verification analysis of 20 GCMs included in the 4[th] IPCC assessment report regarding their representation of the hydrological cycle on the Danube river basin for 1961-2000 and for the 2161-2200 SRESA1B scenario runs. The basin-scale properties of the hydrological cycle are computed by spatially integrating the precipitation, evaporation, and runoff fields using the Voronoi-Thiessen tessellation formalism. The span of the model-simulated mean annual water balances is of the same order of magnitude of the observed Danube discharge of the Delta; the true value is within the range simulated by the models. Some land components seem to have deficiencies since there are cases of violation of water conservation when annual means are considered. The overall performance and the degree of agreement of the GCMs are comparable to those of the RCMs analyzed in a previous work, in spite of the much higher resolution and common nesting of the RCMs. The reanalyses are shown to feature several inconsistencies and cannot be used as a verification benchmark for the hydrological cycle in the Danubian region. In the scenario runs, for basically all models the water balance decreases, whereas its interannual variability increases. Changes in the strength of the hydrological cycle are not consistent among models: it is confirmed that capturing the impact of climate change on the hydrological cycle is not an easy task over land areas. Moreover, in several cases we find that qualitatively different behaviours emerge among the models: the ensemble mean does not represent any sort of *average model*, and often it falls between the models' clusters.

---

[*] Corresponding author, email: lucarini@adgb.df.unibo.it

## 1. Introduction

The Danube river, whose length is approximately 2850 Km, originates from the Black Forest (Germany) and flows into the Black Sea in an extended Delta region shared between Romania and Ukraine. The 1961-2000 average discharge at the entrance of the Delta (Ceatal Izmail station, 45.22°N, 28.73°E) has been about 6500 $m^3s^{-1}$, (Global Runoff Data Center, Germany, http://grdc.bafg.de/). The Danube basin covers about 800,000 $km^2$ and is rather diverse in terms of geography and climatology. The Alps in the west, the Dinaric-Balkan mountain chains in the south and the Carpathian mountain bow receive the highest annual precipitation (1000–3200 mm per year) while the Vienna basin, Pannonian basin, Romanian and Prut low plains, and the lowlands and the Delta region are very dry (350–600 mm per year). The geographical setting of the Danube river basin is shown in Fig. 1.

The Danube has a manifold importance for the history, economy, politics for the European continent. From a climatic point of view, the Danube basin is especially interesting because it is within continental Europe, while featuring a close connection to the Mediterranean region. On one side, the Danube, as it flows into the connected Black sea basin, gives a relevant contribution of freshwater flux (on average, more than twice the Nile's) into the Mediterranean sea. This bears relevance on a global scale, because of the impact of the Mediterranean outflow in Gibraltar on the global oceanic circulation (Rahmstorf, 1998; Artale et al., 2002; Calmanti et al., 2006). On the other side, whereas the influence of the Atlantic climate is strong in the upper basin (Germany), especially in the central and lower basin a relevant contribution to the water balance comes from precipitated water of Mediterranean origin, basically because the basin is downwind of the dominant westerlies and has a complex orography (Speranza, 2002). The seminal ALPEX experiment has emphasized the relevance of the Alps and of the Mediterranean waters in modifying and enhancing the primary storms of Atlantic origin, through the mechanisms of orographic secondary cyclogenesis and latent heat-driven enhancement (Speranza et al, 1985; Tibaldi et al., 1990, Alpert et al., 1996). Several intense precipitative and disastrous floodings events have struck in recent years central Europe, and especially the Danube basin and neighboring areas (Becker and Grunewald, 2003; Stohl and James, 2004).

The Danube basin is shared between 19 countries (Germany, Austria, Slovakia, Hungary, Croatia, Serbia, Montenegro, Romania, Bulgaria, Moldova, Ukraine, Poland, Czech Republic, Switzerland, Italy, Slovenia, Bosnia-Herzegovina, Albania, FYR of Macedonia). Such geopolitical complexity boils up to the most *international* river basin in the world; as a consequence, the Danube is the subject of a large number of international projects, including the UNDP/GEF Danube Regional Project (http://www.undp-drp.org/drp/index.html), and International Commissions, including the International Commission for the Protection of the Danube River (http://www.icpdr.org/), just to name a few.

Given the strategic importance of water as a resource and as a source of risks at social, economical, and environmental level (Anderson et al., 2003; Becker and Grunewald, 2003), and the growing number of applications of the climate models outputs, it is crucial to assess how climate models are able to represent the statistical properties of the hydrological balance of river basins. Moreover, because of the process of latent heat release, biases in the representation of the hydrological balance may in turn strongly effect mesoscale as well as synoptic scale meteorological processes: in a dynamical sense, water is also an active component of the climate system.'

## 1.a. General issues in auditing Climate Models

The evaluation of the accuracy of numerical climate models and the definition of strategies for their improvement are crucial issues in the earth system scientific community. On one side climate models of various degrees of complexity constitute tools of fundamental importance to simulate past climate conditions and future climate scenarios, as well as to test theories related to basic atmospheric, oceanic, and coupled physical processes. On the other side their outputs, especially in the context of future climate projections, are gaining more and more importance in several other fields, such as ecology, economics, engineering, energy, architecture, as well for the process of policy-making at national and international level. In terms of influences at societal level of climate-related findings, the impacts of the 4[th] assessment report of the Intergovernmental Panel on Climate Change (IPCCAR4) are definitely unprecedented (IPCC 2007), up to the point that the IPCC has been awarded the 2007 Nobel Prize for Peace. The possibilities of analyzing climate models have recently greatly increased as more and more research initiatives are providing open access to the outputs of climate models, as in the case of the EU 5[th] Framework Programme project PRUDENCE (http://dmi.prudence.dk) for several regional climate models, and of the Project for Climate Model Diagnostics and Intercomparison (PCMDI/CMIP3: http://www-pcmdi.llnl.gov), for the global climate models included in IPCCAR4.

The auditing – to be intended as the overall evaluation of accuracy - of a set of climate models consists of two related, albeit distinct procedures. The first procedure is the *intercomparison*, which aims at assessing the consistency of the models in the simulation of certain physical phenomena over a certain time frame. The second procedure is the *verification*, which boils down to comparing the diagnostics of the models to some corresponding observed (or quasi-observed) quantities.

The procedures of auditing climate models are far from being obvious in epistemological terms also when the simulation aims at reconstructing the past climate, let alone in the case of future projections (Lucarini, 2002). The presence of issues typical of *non-normal sciences*, which do not have a paradigm such as the Galilean one (Kuhn 1970), is related to the complexity of the natural processes, the wide range of involved spatial and temporal scales (Peixoto and Oort, 1992), the technological difficulties of monitoring in detail the actual state of the climate, as well as the ubiquitous and unavoidable presence of structural model errors, and finally to the impossibility, given the entropic arrow, to repeat experiments. Therefore, the provision of robust metrics (Lucarini et al., 2007a) able to assess at least empirically the models' performances on specific physical processes as well as on balances involving conservation principles is becoming more and more a crucial issue in the climate community, as recently evidenced by the explicit *call for ideas* for models metrics made by PCMDI/CMIP3. The need for improving the diagnostics tools used in the investigation of geophysical problems has been recently emphasized in the "20 Years of Nonlinear Dynamics in Geosciences" conference held on June 11-16, 2006 in Rhodes, Greece (Elsner et al., 2007, Tsonis and Elsner, 2007).

In order to provide a synthetic and comprehensive picture of the output of a growing number of climate models, recently it has become common to consider the ensemble mean and the ensemble spread of the model, taken respectively as the arithmetic mean and the standard deviation of the models' outputs for the considered metric. These concepts should be used with great care, because neither the word ensemble, nor the word mean or standard deviation can be used thoughtlessly. Whereas it is always possible to compute the (arithmetic or weighted) mean or standard deviation of a set of numbers, the idea of associating such statistical estimators to something meaningful for the considered set – the mean approximating the truth, the standard deviation describing the uncertainty - relies on the more or less hidden assumptions that a) the set is a probabilistic ensemble, which is formed by equivalent realizations of a given process, and b) that the underlying probability distribution is unimodal. In the prototypical case of a particle in a quadratic potential well with random noise, the probability distribution of the position is Gaussian, which is completely determined by the mean and the standard deviation. The mean is the most probable state, coincides with the minimum of the well and thus describes the deterministic part of the dynamics (the *true* value), the standard deviation measures the effects of the stochastic processes. The closer we are to this case, the more meaningful is to compute the mean and the standard deviation.

We have no theorem nor evidence that the outputs of the various climate models can be considered in any real sense a probabilistic ensemble – there are virtually infinite reasons why the models are not equivalent, and, whereas *it is for sure not correct to average them with equal weight,* *it not sure whether a correct way of weighting them in terms of their overall quality can be defined at all.* Nevertheless, at empirical level, this latter option seems more promising. Moreover, we have no obvious a-priori expectation that the *probability distribution* of a given output of the various models is unimodal, so that unless we verify that this is the case, the mean and the standard deviation are of limited –if any – utility. Even more: they may be misleading. Finally, even neglecting the theoretical issues, we may have a problem of under-sampling, because, *e.g.* in the case of global climate models, we have only one or two tens of members of the *ensemble*.

**1.b. Difficulties in auditing the hydrological cycle of a climate model**

In nature, the 2D+1D (space and time) fields of main interest for evaluating the hydrological balance are characterized by complex statistical properties, since precipitation features temporal intermittency and spatial multifractal nature (Tessier et al, 1993; Deidda, 1999; 2000; Lovejoy and Scherzer, 2006), evaporation and runoff depend very delicately on the local conditions. Therefore, suitably coarse-grained averaged quantities may be more robust for defining auditing procedure both in terms of intercomparison and verification. Comparisons at *face value* (*i.e.* locally in time and space at the highest possible resolution) of model vs. model and, especially, of model vs. observation can be problematic, since the statistical significance of data decrease as we go towards smaller scales, as shown *e.g.* by Accadia et al. (2003). Nevertheless, the procedure of verification, as opposed to the model intercomparison, faces some serious problems if we consider actual observations of precipitation, evaporation, and runoff. Spatial averages of the water balance for the area of interest cannot be determined with reliability from the scattered time series measurements of the surface characteristics. Since rigorous water-conserving gridding of the climatology of observed precipitations is virtually impossible (Lovejoy and Scherzer, 2006), the verification process in the case of river basins is better defined when relying on an integrated quantity such as the sea-discharge of the river. When climatological time scales are considered, the basin integrated value of the difference between precipitation and evaporation must be equal to the basin integrated value of the surface and subsurface runoff (apart from the relatively small infiltrations in the aquifers), and both must be equal to the river discharge at the end of its course, because water is conserved. The average river discharge is also equal to the convergence of water vapor in the atmosphere over the basin, which is strongly constrained by large-scale meteorological processes. This emphasizes the need for framing the hydrological cycle in meteorological rather than in purely geographical terms.

In recent years, several studies focusing on the intercomparison and verification of the hydro-meteorological characteristics and hydrological cycle over specific basins and on various time scales using using GCMs, RCMs, and the ERA40 (Simmons and Gibson 2000) and the NCEP-NCAR reanalysis (Kistler et al. 2001), have been performed. In this regard, we may mention the papers by Roads et al. (1994), Lau et al. (1996), Gutowski et al. (1997), Betts et al. (1998, 1999), Roads and Betts (1999), Hagemann et al. (2004, 2005), Hirschi et al. (2006). In these analyses, the hydrological cycle of large scale basins has been analyzed with a combination of techniques such as the evaluation of the convergence of atmospheric water vapor, precipitation minus evaporation, runoff, and variation of terrestrial water. The auditing procedure of the representation of the hydrological cycle in GCMs has underlined that flaws or discrepancies are caused both by issues in the large scale atmospheric circulation patters affecting the water vapor transport, by biases in basic catalyzing mechanisms (e.g. misrepresentation of the orography), and by problems in the model representation of some severely parameterized processes occurring in the atmosphere, (e.g. hydrometeor formation and precipitation), at the surface-atmosphere interface, (e.g. evaporation), and inside the soil (e.g. water transport).

**1.c. Previous results on the Hydrological cycle of the Danube basin**

In a recent study for the project HYDROCARE (http://www.hydrocare-cadses.net), the authors have audited several regional climate models (RCMs), running on a domain centered on Europe and nested into the same run of the same Atmospheric Global Circulation Model (AGCM), on their representation of the hydrological cycle of the Danube basin for 1961-1990 (Lucarini et al., 2007b). Large discrepancies – larger than 50% - have been found among RCMs for the monthly climatology as well as for the mean and variability of the basic-integrated annual water balances (and accumulated evaporation and precipitation), and only few datasets are consistent with the observed discharge values of the Danube at its Delta, even if the driving AGCM provides itself an excellent estimate. Thus, RCMs seem to degrade the information provided by the large scale flow, once the local, downscaled information they produce is up-scaled to an intermediate range between the minimum resolvable scale and the domain size. Changes in the resolution seem not to be effective for solving this problem, as for a given model increases in the resolution do not alter the net water balance, while speeding up the hydrological cycle through the enhancement of both precipitation *and* evaporation by the same amount. Moreover, some deficiencies of the land models have been evidenced by the fact that for some models the hydrological balance estimates obtained with the runoff fields do not agree with those obtained via precipitation and evaporation. These results pose doubts regarding the ability of most RCMs in representing correctly the hydrological cycle in river basins, because, apart from the structural uncertainties of the model, the biases due to the nesting procedure introduced in the buffer zone seem to be critical (Lucarini et al., 2007b). Another significant result is that the ERA40 and the NCEP-NCAR reanalysis, which are routinely used at face value as verification benchmark also for precipitation, even if deficiencies have been discussed in the literature (Betts et al., 2003; Troccoli and Kallberg 2004; Amenu and Kumar 2005; Hagemann et al., 2005), result to be largely inadequate for representing the hydrology of the Danube river basin − and probably not only there -, thus providing estimates that are not only quantitatively wrong, but qualitatively unreasonable (Lucarini et al., 2007b).

**1.d. This paper**

In this paper we analyze the performances of 20 GCMs participating to the IPCCAR4 in the representation of the hydrological cycle of the Danube basin for the last 40 years of the XX century. Moreover, the changes in the hydrological cycle foreseen by the climate projections in the last 40 years of the XXII century under the SRESA1B scenario (720 ppm of $CO_2$ after 2100) are analyzed. This latter effort is also motivated by the acknowledgment that the Danube basin sits between two "hot-spots" in terms of amplified projected climate change, the so-called Mediterranean and Northern Europe (which nevertheless includes also central Europe) regions (Giorgi, 2006). The hydrological cycle of these two regions seem to respond rather differently under climate conditions: the Mediterranean region should experience a decrease in the average precipitation, whereas the opposite holds for Northern Europe. The SREASA1B scenario has been selected because it is somewhat median within the IPCC scenarios in terms of GHG forcings (IPCC, 2007). For all datasets we compute in two independent ways the long-term averages and the 95% confidence interval of the annual averages. We also show and discuss critically why for several diagnostics the concept of ensemble mean is of limited use.

The paper is structured as follows. In Section 2, basic information about the Danube basin, the data considered in this study and the concepts behind the diagnostics tools employed in the auditing are presented. In Section 3 and 4 we present and discuss the main results on the intercomparison and verification of the models, regarding the yearly climatology precipitation, evaporation, water balance and runoff as given by the simulations relative to 1961-2000 and 2161-2200 for the SRESA1B scenario, respectively. In Section 5 we draw our conclusions.

## 2. Data and methods

**2.a Datasets**

The following data sources relative to the 1961-2000 time-frame have been used for the purpose of our analysis:

1.  Monthly values of Runoff (R), Precipitation (P), Evaporation (E), from 20 GCMs collected from the PCMDI/CMIP3 data portal (http://www-pcmdi.llnl.gov/), containing all simulations relevant for the compilation of the IPCCAR4 - see Table1.

2.  Monthly values of R, P, and E from the 2 major reanalysis datasets, ERA40 and NCEP-NCAR - see Table 1. For the latter reanalysis, E data has been obtained straightforwardly from the Latent Heat Flux data. The data have been downloaded from http://data.ecmwf.int/data/d/era40_daily/ (ERA40) and http://www.cdc.noaa.gov/cdc/reanalysis/reanalysis.shtml (NCEP-NCAR)

3.  Monthly discharge (D) of the Danube river at the Ceatal Izmail station (45.22°N, 28.73°E), which is the last one before the Delta. Data have been obtained from the Global Runoff Data Center (GRDC), Germany (http://grdc.bafg.de/).

4.  Monthly values of R, P, and E from the 20 GCMs as in 1. , for the 2161-2200 simulations under the SRESA1B scenario, featuring a stable 720 ppm $CO_2$ concentration after 2100.

**2.b Theoretical framework**

By imposing instantaneous mass conservation for water and assuming that water storage is negligible when long-term averages are considered (Peixoto and Oort, 1992), at any point in space the difference between P and E equals R, which also equals the horizontal convergence of the vertically integrated atmospheric water flux ($\vec{Q}$):

$$\langle P \rangle_\tau - \langle E \rangle_\tau \approx \langle R \rangle_\tau \approx -\langle \vec{\nabla}_H \cdot \vec{Q} \rangle_\tau \qquad (1)$$

where $\langle \bullet \rangle_\tau$ indicates the operation of averaging over time $\tau$. This equation is valid on time scales $\tau \geq 1\,y$, which are long compared to the average residence time of water in the atmosphere ($\sim 10$ days), to the duration of the temporary storage of water in form of snow cover ($\sim$ few months at most), and to the seasonal cycle (3-6 months). If we integrate spatially Eq. (1) over the entire geographical region $A$ corresponding to the hydrological basin of a river, we obtain the following basic form of hydrological balance:

$$\int_A dx dy \left( \langle P \rangle_\tau - \langle E \rangle_\tau \right) = \int_A dx dy \langle B \rangle_\tau \approx -\int_A dx dy \langle \vec{\nabla}_H \cdot \vec{Q} \rangle_\tau \approx \int_A dx dy \langle R \rangle_\tau \approx \langle D \rangle_\tau \qquad (2)$$

where $B = P - E$ is the net balance and $D$ is the actual river discharge into the sea. Note that, considering the Gauss integral theorem, we also have that the time-averaged river discharge equals the time average of the net incoming atmospheric flux of water through the vertical boundaries of the atmospheric region bounded below by the region $A$: in a way, the river flows down from the sky. Equation (2) forms the basis of the diagnostic study presented in this paper. For each model we define the yearly time series of the accumulated basin integrated fields as follows:

$$\overline{\Phi}_i \equiv \int_A dx dy \langle \Phi \rangle_i \,, \qquad (3)$$

where $\Phi$ is any of the field $P, E, B,$ or $R$, A is the geographical domain of the basin, and the time averaging on the right hand side of Equation (3) is performed over the calendar year indicated as the lower index,. Therefore, $\overline{\Phi}_i$ is a time series composed of 40 elements. Using standard statistics, we have that the best unbiased estimate of $\overline{\Phi}_i$ can be written as:

$$\mu(\overline{\Phi}_i) = \frac{1}{40} \sum_{i=1}^{40} \overline{\Phi}_i \,, \quad (4)$$

while the best unbiased estimate for the standard deviation of $\overline{\Phi}_i$ can be written as:

$$\sigma(\overline{\Phi}_i) = \left[ \frac{1}{39} \sum_{i=1}^{40} \left( \overline{\Phi}_i - \mu(\overline{\Phi}_i) \right)^2 \right]^{1/2} \,, \quad (4)$$

Assessing the mutual consistency between the climatologies provided by the various models entails comparing their estimates of $\mu(\overline{\Phi}_i)$, $\sigma(\overline{\Phi}_i)$, *and* considering the statistical uncertainties associated with such estimates. For all models, for both the 1961-2000 and 2161-2200 time frames, and for all fields $\Phi$, the time series $\overline{\Phi}_i$ do not have any statistically significant trend and are compatible with the null hypotheses of Gaussian white noise. In all cases considered, the estimates of the absolute value of the lagged correlations are smaller than 0.25 for all time lags $\geq 1$ year, whereas the corresponding 95% confidence interval for a synthetic white noise time series of the same length is about [-0.32, 0.32]. Therefore, it makes sense to define $\left[ \mu(\overline{\Phi}_i) - \delta(\mu(\overline{\Phi}_i)), \mu(\overline{\Phi}_i) + \delta(\mu(\overline{\Phi}_i)) \right]$ as the 95% confidence interval of $\mu(\overline{\Phi}_i)$, and $\left[ \sigma(\overline{\Phi}_i) - \delta(\sigma(\overline{\Phi}_i)), \sigma(\overline{\Phi}_i) + \delta(\sigma(\overline{\Phi}_i)) \right]$ as the 95% confidence interval of $\sigma\left(\overline{\Phi}_i\right)$. Using the gaussian white noise approximation, it is possible to express the uncertainties on the estimates as simple functions of the standard deviation of the time series, so that $\delta\left(\mu\left(\overline{B_i}\right)\right) \approx 2\,\sigma\left(\overline{B_i}\right)/\sqrt{N}$ and $\delta\left(\sigma\left(\overline{B_i}\right)\right) \approx \sqrt{2}\,\sigma\left(\overline{B_i}\right)/\sqrt{N}$ with $N=40$. The approximate equalities become exact in the asymptotic limit. Similar (within 10%) widths of the confidence intervals are obtained through various methods, such as block-bootstrap resampling, which takes into more detailed account time-lagged correlations (Wilks, 1997; Lucarini et al., 2006).

## 2.c Data manipulation

In order to obtain the yearly time-series of the basin-integrated values of the precipitation, evaporation, and runoff fields, MATLAB 7.0.4®, MS Office Excel® and ArcGIS 9.0® software packages have been used with customized routines. After data manipulation, a GIS point layer of all characteristics, including the actual boundaries of the basin, is created. The numerical integration of the field has been performed using Voronoi or Thiessen tessellation (Okabe et al., 2000). The strategy can be described as follows. We first define for each grid point the corresponding grid cell, defined as the set of points that are closer (in the natural metric) to that grid-point than to any of the other ones. Then, we assume that the field under consideration evaluated at a grid-point is constant within the corresponding grid-cell. Then, all the contributions coming from cells contained inside the geographical domain of the river basin are summed up with a weight corresponding to the area of the cell, while the contributions from the boundary cell (not entirely contained in the basin) are weighted with the portion of the area of the cell inside the basin. See Figs. 2a-2c for some examples. Using this approach, we introduce no spurious information, as always done when interpolating the gridded data. Moreover, we are guaranteed that no water is lost or added in the upscaling. In the cases considered in this study, since the grids are locally quasi-rectangular in the natural metric, the Voronoi tessellation is such that the grid cell corresponding to the grid-point $(j,k)$ is basically a rectangle with corners given by the combinations of the grid points $[(j\pm1,k\pm1)+(j,k)]/2$. See Figs. 1b-1d. It has to be noted that the actual geographical outline of the Danube basin does not necessarily correspond to how the models individually represent it, since the GCMs are bound to their spatial resolution and representation of orography. Nevertheless, we believe that our procedure provides a robust common strategy valid for all models when bulk, integrated properties of the hydrological cycle, as the considered basin area – and so the effective integration domain - is the same for all models.

## 3. Results: Present Climate

### 3.a. Water Balance

We start by considering $B = P - E$. Results are presented in Fig. 3. The scatter plot portrays the 95% confidence interval of the best estimate of the interannual variability $\sigma(\overline{B}_i)$ of the integrated water balance for each of the 20 GCMs considered in this study. Figure 3 displays that on one side basically all GCMs agree on the interannual variability of the water balance, which is around 70 $mm$ $y^{-1}$, with the exception of the CGCM3T63 model, which features a significantly lower variability. When considering the estimates of the yearly averages, the agreement between models is much lower: the span of the outputs is as large as the observed balance, with only four models – ECHOG, GISSAOM, UKMOHADCM3 and CGCMT47 in statistical agreement with the observed climatological discharge of the Danube. The model outputs are not clearly clustered, except for a somewhat distinct group of models – PCM1MODEL, UKMOHADGEM1, ECHAM5, and GISSEH – which feature a serious dry bias, nor there is a notable positive correlation between the mean yearly water balance and its interannual variability. Moreover, a higher resolution model does not guarantee a better agreement with data. Note that, whereas the two versions of the MIROC model differing for the resolution have an almost identical water balance, in agreement with what observed in Lucarini et al. (2007b), this is not the case for the two CGCM models, where the water balance given by the T63 model is statistically not compatible with what given by the T47 model. A serious error we have found is one negative entry for yearly-averaged, basin integrated water balance given by the GISSER model. This is a physically unreasonable result, since it would imply that the Danube basin is a net exporter of water, so that the presence of a river would be impossible. In general, in order to have at the same time a long-term negative surface water balance and the actual presence of a river, large amounts of groundwater have to flow upward. Nevertheless, this does not apply to this situation, given the characteristics of the model hydrology, and, in any case, the large size of the region involved.

For reference, the figure portrays also the GCMs ensemble mean – computed with the usual arithmetic averaging - for the yearly water balance. The value of the ensemble mean falls in a range of values *populated* by models and its agreement with observations is comparable to that of the 6 best models. In this case, the adoption of such a proxy for the overall performance of the set of GCMs seems fine, but we will see that considering different diagnostics things become more troublesome.

As already observed in Lucarini et al. (2007b) for the 1961-1990 time frame, the ERA40 reanalysis gives a 95% confidence interval for the yearly water balance which intersects 0, resulting from several negative entries of the yearly time series. The NCEP–NCAR best estimate of the yearly water balance is about half of observations, with an interannual variability which is about twice of that of ERA40 and of all GCMs. A novel result with respect to what presented in Lucarini et al. (2007b) is that also the NCEP reanalysis gives some negative entries of the yearly water balance time series in the last decade of the last century. These results call for a more consistent treatment of the water vapor in the reanalyses, otherwise they cannot be considered as a benchmark for GCM performance of precipitation, evaporation, and water balance.

**3.b. Strength of the Hydrological cycle, Precipitation, and Evaporation**

In Fig. 4 we represent the statistical properties of the time series of the yearly averaged basin integrated strength of the hydrological cycle $H = P + E$ for 1961-2000. Whereas there is an overall agreement among models regarding the interannual variability of $\overline{H}_i$, a very large spread is observed for of the average strength of the hydrological cycle, with the GCMs outputs ranging from about 1050 $mm\ y^{-1}$ to about 1650 $mm\ y^{-1}$. Since the width of the typical 95% confidence interval of $\hat{\mu}\left(\overline{H}_i\right)$ is around 50 $mm\ y^{-1}$, we have that the overall statistical agreement among models is modest but still notably better than those of the reanalyses. Moreover, models tend to cluster into two statistically well-distinct groups, one characterized by a low value of $\mu\left(\overline{H}_i\right)$ (between 1050 $mm\ y^{-1}$ and 1200 $mm\ y^{-1}$), the other one characterized by higher values of $\mu\left(\overline{H}_i\right)$ (between 1350 $mm\ y^{-1}$ and 1650 $mm\ y^{-1}$), so that the ensemble mean falls between the two clusters. In such a situation, where the distribution of the model outputs – taken as samples of a probability space - does not resemble anything like a unimodal distribution, we maintain that the ensemble mean contains a rather modest information on the actual outputs of the GCMs.

Similarly to what found in Lucarini et al. (2007b), we have that GCMs having rather similar water balance feature widely different intensities – up to a factor of 1.5 - of the hydrological cycle. Clear examples of this – and in general of the difficulty in classifying GCMs in terms of a single metric – are, among the *dry* models, PCM1MODEL and UKMOHADCM3, and among the *wet* models, FGOALS and INMCM30. This implies that GCMs greatly differ among each other in the ratio between basin-integrated yearly accumulated precipitation and evaporation, so that it is not possible to use the climatological properties of the precipitation as a quick-and-dirty proxy of the climatological properties of the whole hydrological cycle.

The poor performance of the two renalyses is demonstrated by the fact that the discrepancy between their statistical properties shown in Fig. 4 is larger than the discrepancy between any pair of GCMs considered and no overlap exists.

Further information on the skill of the considered datasets can be obtained by analyzing the basin-integrated yearly-accumulated time series of precipitation and evaporation. Results are presented in Fig. 5. We first observe that the width of the 95% confidence interval of $\mu(\overline{P}_i)$ is in all cases much larger, by almost an order of magnitude, than that of $\mu(\overline{E}_i)$, thus implying that precipitation has a much larger interannual variability.

In the case of MIROC models, the version with higher resolution features an enhanced precipitation and a compensating enhanced evaporation, so that the water balance in unchanged. Basically, what happens is that the large scale atmospheric water influx into the basin and the efficiency of large scale precipitation are not changed, whereas the local processes, responsible for speeding up the hydrological cycle, become stronger with increased resolution. This is analogous to what found when comparing various versions of RCMs nested in the same run of the same GCM (Lucarini et al 2007b). In the case of the two CGCM models, the change in the resolution affects the water balance, which implies a disagreement in the large scale features of water transport and in the impact of differently resolved orography on precipitative processes.

We next look at the joint statistical properties of the $\overline{P}_i$ and $\overline{E}_i$ time series. Considering that the 95% confidence level for the null hypothesis of absence of correlation is about 0.32, in most of the GCMs it is found a high degree of positive correlation between the two time series, which is the signature of the precipitation-evaporation positive feedback. The following mechanism is in place on a local scale: higher precipitation leads to a moister soil, and then to a higher evaporation, which eventually increases the water content of the atmosphere. The values of the correlation coefficients are notably lower than what obtained with RCMs (Lucarini et al. 2007b), which is consistent with the fact that in RCMs, which have higher resolution, small scale precipitative (and evaporative) features have a greater importance. Moreover, in any case of four GCMs – MIROChr, MIROCmr, CGCM3T63, MRICGCM, ECHOG - the correlation between $\overline{P}_i$ and $\overline{E}_i$ is not statistically significant, which may suggest that in these models the interaction between the atmosphere and the underlying soil is somewhat weaker.

When considering the two reanalyses, we obtain rather similar results to those discussed in Lucarini et al. (2007b). They completely disagree regarding the intensity of the hydrological cycle, which is rather strong for NCEP-NCAR and rather weak for ERA40, and they have opposite behaviors regarding the precipitation-evaporation feedback, with the NCEP-NCAR dataset showing a rather robust positive correlation between $\overline{P}_i$ and $\overline{E}_i$, whereas for ERA40 the correlation is weakly negative but not statistically significant. This suggests that the local precipitative processes do not have an efficient parameterization in the model underlying the ECMWF reanalysis system (Troccoli and Kallberg 2004; Amenu and Kumar, 2005).

### 3.c. Runoff

We can further assess the quality of the outputs of the GCMs by making a consistency check obtained by comparing for each model. We then verify how precisely models conserve water by comparing the time-averaged water balance obtained by integrating the difference between the precipitation and the evaporation fields to the basin-integrated runoff. The runoff is of the uttermost importance when the results of climate simulations are used as input of agricultural, societal, and environmental models. In Fig. 6 we present the 95% confidence interval of $\mu\left(\overline{R}_i\right)$ plotted against the 95% confidence interval of $\mu\left(\overline{B}_i\right)$. The observative constraint, *i.e.* the mean discharge at the Ceatal Izmail station, is also indicated.

We see that for most GCMs the climatological basin-integrated estimates of the water balance and of the runoff agree within statistical uncertainty. Nevertheless, in several cases the agreement is somewhat marginal and the best estimate is below the bisectrix. In four cases - INCM30, IPSLCM4, UKMOHADGEM, GISSER – the climatological runoff estimates are definitely larger than those of the water balance by about 10%-15%, thus implying that, on the average, water is created in the soil components of the models. A very large disagreement – of the order of 40% - is detected in the case of the FGOALS model, where, on the contrary, the code defining the runoff *loses* large amounts of water.

The fact that the agreement between the climatology of the water balance and of the runoff on such long time-scales is far from being perfect in most models - the main exceptions being ECHAM5, PCM1MODEL, MIROChr – calls for a detailed audit of the soil components of the models. Regarding the two reanalyses, the issues evidenced in Lucarini et al. (2007b) are confirmed. The soil modules of NCEP-NCAR and ERA40 create on the average an amount of water corresponding to half of the Danube discharge and to the whole Danube discharge, respectively.

It is reasonable to expect that the integrated runoff should have intrinsically a smaller variability because it results from a sequence of processes occurring in the soil and acting effectively as a low-pass filter on the instantaneous water balance. This picture is basically confirmed in all models, as the width of the 95% confidence interval for $\mu\left(\overline{R}_i\right)$ is typically about 15% smaller than that of $\mu\left(\overline{B}_i\right)$.

## 4. Results: Impact of Climate Change

In this section we analyze the impact of the change of atmospheric composition foreseen by the SRESA1B scenario (fixed $CO_2$ concentration of 720 ppm after 2100) on the hydrological cycle of the Danube basin. We have selected the end of the XXII century instead of the end of XXI century in order to have all the transient effects due to the increase in the $CO_2$ to die out. As mentioned before, the time series we analyze do not feature any statistically significant trend. Unfortunately, not all GCMs considered in the previous section provide data for the 2161-2200 time frame, so that we restrict our analysis to only 16 of the 20 models. A still smaller group of models provide data for the end of the XXIII century; we have consistently verified that the statistical properties of the hydrological cycle for the 2261-2300 time frame agree with those of the 2161-2200 time frame, thus further confirming that by the end of XXII century stationary climatological properties are realized.

For each model, the change in the statistical properties between the 1961-2000 and the 2161-2200 period is assessed by computing the 95% confidence interval of the variation of the mean - $\Delta\mu(\overline{\Phi}_i) = \mu(\overline{\Phi}_i)^F - \mu(\overline{\Phi}_i)^P$ - and of the standard deviation - $\Delta\sigma(\overline{\Phi}_i) = \sigma(\overline{\Phi}_i)^F - \sigma(\overline{\Phi}_i)^P$ - of the time series $\Phi_i$, where the superscripts $F$ and $P$ refer to the 2161-2200 and 1961-2000 time frames, respectively. Since the two time series are uncorrelated, the width of the confidence intervals of $\Delta\mu(\overline{\Phi}_i)$ ($\Delta\sigma(\overline{\Phi}_i)$) is computed by summing in quadrature the widths of the confidence intervals of $\mu(\overline{\Phi}_i)$ ($\sigma(\overline{\Phi}_i)$) obtained in the two time frames.

### 4.a. Water balance

Fig. 7 shows the change between the statistics of the 2161-2200 and the 1961-2000 basin-integrated annual accumulated water balance. We plot the 95% confidence interval of the change in the mean, $\Delta\mu(\overline{B}_i)$, vs. the 95% confidence interval of the change in the interannual variability, $\Delta\sigma(\overline{B}_i)$. Apart from UKMOHADCM3, all models show a negative change for the mean. For CSIRO and PCM1MODEL the changes are not significant (*i.e.* error bars intersect the zero line). The region 3 of Fig. 7 is empty, except for the dot representative of UKMOHADCM3. Nevertheless, for this model the response of the water balance of the Danube basin to the global climate change is rather weak, so that no statistically significant changes of both moments of the distribution can be detected. This is surely a matter calling for a detailed investigation. Most models show also a tendency for decreasing the interannual variability (region 2 is quite populated), even if for only one model (GISSER) such change is statistically significant. Finally, few models show an increase in the interannual variability of the balance (region 1), even if for only one - CGCMT63 – the change is statistically significant. It is interesting to note that CGCMT47 and CGCMT63 feature rather different responses to climate change. We have found further negative entries for the yearly averaged basin integrated water balances of GISSER, CNRMCM, and UKMOHADGEM. The fact that the number of physically unreasonable balances greatly increase and affects three models instead of one suggests that the flaws in the representation of the water exchanges between atmosphere and soil are evidenced when conditions of increased surface temperature are realized. This points at a misrepresentation of the evaporative processes.

Changes in the regime of water balance can be evidenced by computing the variation between 1961-2000 and 2161-2200 of the normalized interannual variability, taken as the ratio between the actual interannual variability and the average yearly water balance. This quantity measures how stable is the yearly water balance and is related to the probability of having entire years characterized by serious water scarcity in the entire basin. Figure 8 shows that for most GCMs the normalized interannual variability increases in statistically significant way, so that the relative fluctuations of the yearly water balance become larger. This is qualitatively consistent with the results shown in Giorgi and Bi (2005a), where it is shown that the increase of the ratio between the interannual variability and the mean value of the precipitation is a large scale feature in Europe and beyond.

In particular, for some models (e.g. ECHOG, ECHAM5, CNRCM), the stability of the water balance is greatly decreased under climate change condition, since the normalized interannual variability almost doubles. The models which have been found to feature qualitative deficiencies in the treatment of the water balance (GISSER, CNRMCM, UKMOHADGEM) have the largest normalized interannual variability for 2161-2200, with GISSER featuring also the largest sensitivity to climate change. Instead, for INMCM30, FGOALS, UKMOHADCM3, featuring a very stable water balance with low normalized interannual variability, and for PCM1MODEL, having opposite characteristics, the effect of climate change is negligible.

When assessing for each model the self-consistency between the long-term integrated difference between precipitation and evaporation and the integrated runoff for the XXII century simulations, results basically confirm what obtained for the XX century simulations. This reinforces the idea that the biases described in section 3.c depend basically on the model structure rather than on the specific simulations considered.

**4.b. Strength of the Hydrological cycle, Precipitation, Evaporation**

In order to interpret the change in the water balance of the GCMs, we now look into the changes of the yearly accumulated basin integrated precipitation ($\Delta\mu(\overline{P_i})$) and evaporation ($\Delta\mu(\overline{E_i})$) between the reference period and the 2161-2200 time frame. Results are depicted in Fig. 9. Since for all GCMs (except UKMOHADCM3) the mean water balance $\mu\left(\overline{B}_i\right)$ decreases, the difference between the change in the precipitation and the change in the evaporation is negative, so that the representative dots lie on the left hand side of the bisectrix and the region 4 is empty. It is interesting to note that the dots are split into roughly even parts between the three regions 1, 2, and 3. Models in region 1 feature a decrease in precipitation which is accompanied by a smaller decrease in evaporation. Models in region 2 feature, on the other side an increase in the precipitation, together with a larger increase in evaporation. The consistency between models in capturing the changes in the mean precipitation is very poor: this can be attributed to the fact that models tend to give opposite climate change signals for precipitations in the previously mentioned "hot-spots" regions sharing geographically the Danubian basin, the Mediterranean region and the Northern Europe region (Giorgi and Bi, 2005a; 2005b; Giorgi, 2006). Such *boundariness* of the Danubian basin is also evidenced by the fact that the correlations of the local precipitations with the North Atlantic Oscillation (NAO) signal changes sign within the basin, so that, as discussed in Lucarini et al. (2007b), the basin averaged precipitation (and evaporation) signals have no correlation with NAO.

Since in the 2161-2200 climatic conditions the evaporation over wet surfaces is enhanced due to the larger capacity of a warmer atmosphere to retain water vapour, the behaviour of GCMs belonging to regions 1 and 2 implies that for these models the evaporative process tends to be water-limited, with the difference that in the two regions the change of the availability of water has opposite signs. For models belonging to region 3, instead, we have a decrease of the precipitation accompanied by an increase of the evaporation, so that the evaporation is poorly constrained by the limitation in the water availability. Therefore, there is a qualitative difference between the behaviour of models belonging to regions 1 and 2 with respect to those belonging to region 3. The ensemble mean of the GCMs cannot really represent any sort of *overall average model*.

Held and Soden (2006) found that, in the geographical regions where liquid water is always available at surface (*e.g.* the ocean), in global warming conditions the change of the water balance is proportional to the present climate value, *i.e.* the water balance increases where it is positive and decreases where it is negative. Over land regions, where water is limited, this statement may not apply. Consistently, we find a break-down of the Held and Soden picture, as in the Danubian region $\Delta\mu\left(\overline{B}_i\right)$ is negative whereas the present value of $\mu\left(\overline{B}_i\right)$ is positive.

Another signature of the fact that the estimate of the changes induced by global warming of the hydrological cycle in the land areas is not robust among models – as to be expected given the complexity of the soil-atmosphere interaction and the consequent sketchiness of its parameterizations - is provided by Fig. 10, where we present a scatter plot of a measure of the yearly averaged basin integrated strength of the hydrological cycle $\mu(\overline{H_i}) = \mu(\overline{P_i}) + \mu(\overline{E_i})$ of the 1961-2000 versus the 2161-2200 period. Four models (INMCM30, UKMOHADCM3, UKMOHADGEM, FGOALS) do not foresee any significant change in the intensity of the hydrological cycle, 5 models (ECHAM5, GISSER, IPSLCM4, CNRMCM, GFDL20) foresee a decrease of the intensity of the hydrological cycle, and the remaining 7 models go to the opposite direction. The three categories here evidenced include member of both clusters evidenced in Fig. 4, so that the intensity of the hydrological cycle in the present climate conditions does not discriminate. Of course, GCMs belonging to region 1 (2) in Fig. 9 feature an decrease (increase) in the strength of the hydrological cycle, whereas those belonging to region 3 tend to have small changes. Given the qualitatively distinct behavior of the GCMs, any averaging performed assuming they are all equivalent (like done in ensemble mean) does not provide any really useful information.

We would like to emphasize an indirect signature of the change of the nature of precipitations in changed climate conditions characterized by global warming, going back to the joint statistical properties of the $\overline{P_i}$ and $\overline{E_i}$ time series presented in Table 2. Basically for all models the correlation is higher for the 2161-2200 time frame than for the 1961-2000 time frame, so that the number of GCMs featuring a statistically significant positive correlation between the time series in increased. This implies that in the foreseen climate conditions the link between precipitations and evaporation in this area is stronger. An explanation for this behavior is that in warmer condition evaporation of the water contained in the soil is more efficient, and consequently the local precipitative events – which do not alter the water balance - become more relevant relatively to the total precipitation.

## 5. Summary and Conclusions

This paper provides an analysis of the simulations of the hydrological cycle of the Danube basin performed by GCMs in the context of the present climate (1961-2000) and of the future climate projections (2161-2200), as foreseen by the IPCC scenario SRESA1b, which is sort of median in terms of foreseen GHGs concentrations. This work aims at validating GCMs in the representation of climatic fields of great importance both in purely scientific terms and with respect to impacts at socio-environmental level.

The spatial integration within the basin is performed by constructing the Voronoi tessellations of the gridded fields produced by the models, and successively weighting the contribution of each grid-point with the fraction of the area of the corresponding Voronoi cell contained inside the actual geographical boundaries of the Danube basin. Whereas each model gives a different description of the basin, which basically depends on the spatial resolution and on the representation of the orography, the procedure of using for all models the real geographic boundaries seems robust for evaluating bulk properties of the hydrological cycle, basically because in this way the effective integration domain – results to be the same for all models.

Regarding present climate, basically all GCMs agree on the interannual variability of the water balance (around 70 mm y$^{-1}$); for the yearly averages, the agreement between models is much lower: the span of the outputs – from 100 to 400 mm y$^{-1}$ - is as large as the observed balance – about 250 mm y$^{-1}$ - , and only four models – ECHOG, GISSAOM, UKMOHADCM3 and CGCMT47 – are in statistical agreement with the observed climatological discharge of the Danube. Discrepancies in the water balance greatly effect the energy balance of the atmosphere: a discrepancy of 100 mm y$^{-1}$ in the water balance corresponds to a discrepancy of about 7.2 W m$^{-2}$ in the energy balance of the atmosphere, with the ensuing impact on circulation structure through potential vorticity isosurfaces stretching due to (differential) heating. When considering the strength of the hydrological cycle, there is an overall agreement between models regarding the interannual variability, whereas a very large spread is observed for the averages, with the GCM outputs ranging from 1050 mm y$^{-1}$ to 1650 mm y$^{-1}$. The GCMs greatly differ in the ratio between P and E, so that it is not possible to use the climatological properties of the precipitation only as a quick-and-dirty proxy of the properties of the whole hydrological cycle. In most GCMs a high degree of positive correlation is found between the P and E time series, which is the signature of the precipitation-evaporation positive feedback. The observed spread and the degree of statistical agreement of independent runs of various relatively coarse resolution GCMs among themselves is rather similar to what shown in Lucarini et al. (2007b) regarding a set of high-resolution RCMs forced at the boundaries by the same run of the same GCM. This suggests that it is a non-trivial task to construct high-quality climatological statistics with RCMs, basically because the nesting procedure may introduce spurious biases.

The inadequacy of the NCEP-NCAR and ERA40 reanalyses is confirmed as the discrepancy between their statistical properties is larger than that between any pair of GCMs considered.

For basically all GCMs the water balance decreases in the SRESA1B scenario, which confirms the fact that the general findings of Held and Solden (2006), foreseeing a climate change induced variation of the water balance having the same sign as the present water balance, do not strictly apply on land. For some GCMs both P and E increase, with the latter increasing more. For other models, P and E both decrease, with the latter decreasing less. For these models the evaporative process tends to be water-limited. For other models, the decrease in P comes with an increase in E, so that water  the evaporation is poorly constrained by the limitation in the water availability. The interannual variability of the balance increases in relative terms for most models, suggesting that in SRESA1B scenarios water scarcity and drought conditions may become a more serious issue. As opposed to the common wisdom, the response of the GCMs to climate change is not consistent regarding the strength of the hydrological cycle: 4 models do not foresee any change, 5 models foresee a decrease of the intensity of the hydrological cycle, and the remaining 7 models go in the opposite direction. Model-wise, there is no signature of correlation between the sign of the change of the strength of the hydrological cycle and the actual strength in the present climate simulations. Finally, the precipitation-evaporation feedback is reinforced due to increase in local processes. The Danube basin is geographically shared by two regions, the Mediterranean and the Northern European region which are both "hot-spots" for climate change (Giorgi 2006), but which nevertheless differ a lot when foreseen changes in precipitations are considered (Giorgi and Bi, 2005a; 2005b). Our results confirm and emphasize that it is very hard to find agreements among models on the projected changes of the properties of the hydrological cycle of one of the major river systems of Europe. Therefore, a lot of research is needed on regional climate changes and especially on the impacts of climate change on the hydrological cycle at various space-time scales.

We hope that this work, basically a diagnostic and methodological one, may stimulate further analyses aimed at understanding the physical processes determining the specific behaviour of the GCMs analyzed here

We conclude by emphasizing that, in addition to the described qualitatively different behaviours among GCMs, the distribution of model outputs is typically not unimodal. Therefore, the ensemble mean often falls into *nowhere's land*, *i.e.* between the models' clusters, so that the amount of information it contains is very limited. In general terms, this calls for attention in associating to the ensemble mean more reliable estimates of the real state of the system, and we hope that the present study may be stimulating in this direction.

## Acknowledgments

The authors acknowledge the international modelling groups for providing their data for analysis, the PCMDI (http://www-pcmdi.llnl.gov) for collecting and archiving the model data, the JSC/CLIVAR Working Group on Coupled Modeling (WGCM) and their Coupled Model Intercomparison Project (CMIP) for organizing the model data analysis activity, and the IPCC WG1 TSU for technical support. The Data Archive at Lawrence Livermore National Laboratory is supported by the Office of Science, U.S. Department of Energy. The provision of data by the GRDC (http://grdc.bafg.de/) for the monthly Danube discharge is acknowledged. The preparation of the paper has greatly benefited from the discussions with K. Emanuel, F. Giorgi, G. Monacelli. The comments of three anonymous reviewers have been greatly beneficial for the quality of the paper. V.L. gratefully acknowledges the kind hospitality given by ISAC-CNR, Bologna. This work has been performed in the context of the 2006-2007 EU INTERREG IIIB-CADSES project HYDROCARE (http://www.hydrocare-cadses.net).

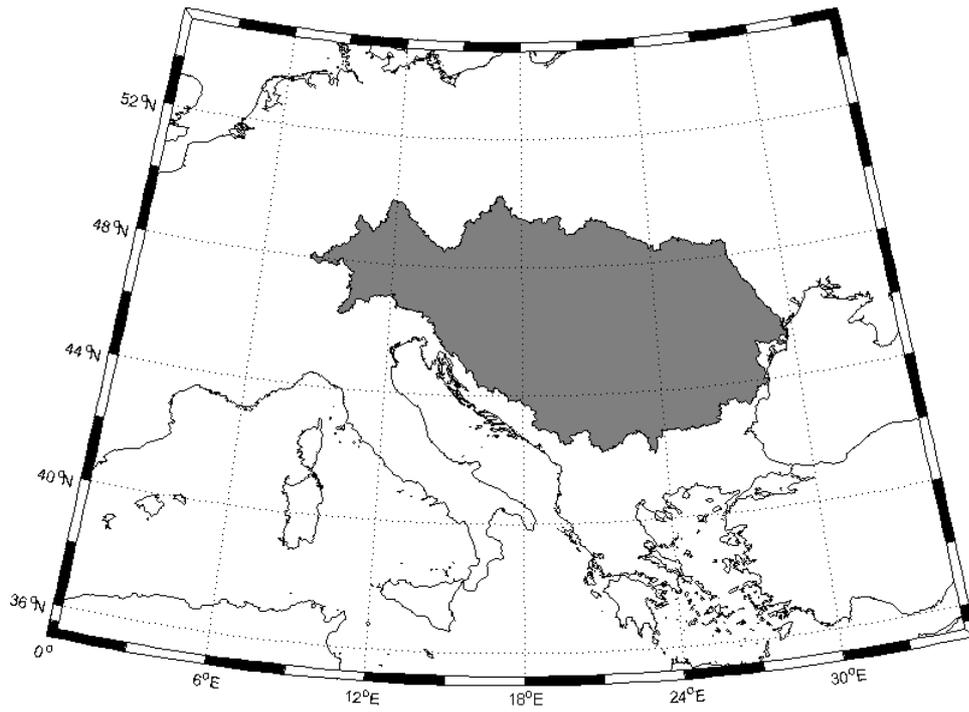

1
2 **Figure 1.** The Danube river basin (shaded in gray) is roughly contained between 42°N and 50°N and between 8°E and
3 29°E.
4

1 **(a)**

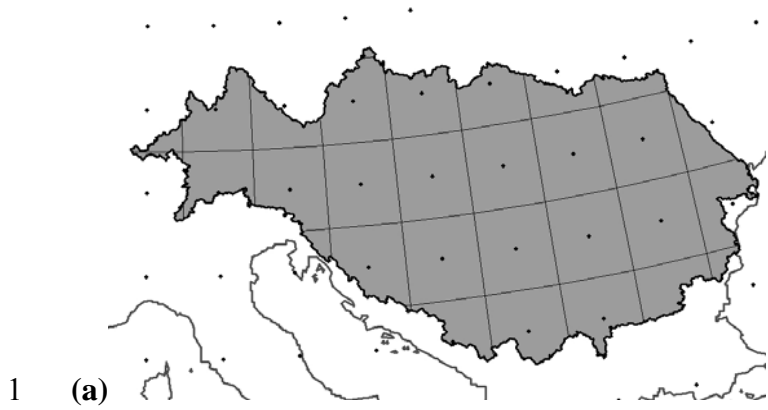

2 **(b)**

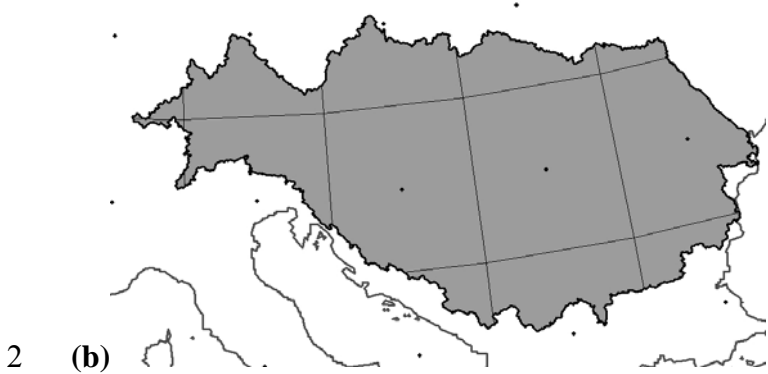

3 **(c)**

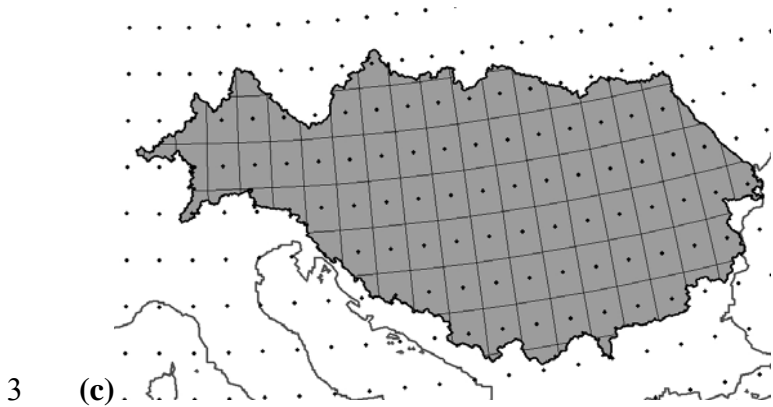

4 **Figure 2.** GIS data processing: point layer to Voronoi polygon layer transformation. The grids of the GFDL2.0 model

5 (a), GISSEH (b), and MIROChr (c) are shown for explanatory purposes.

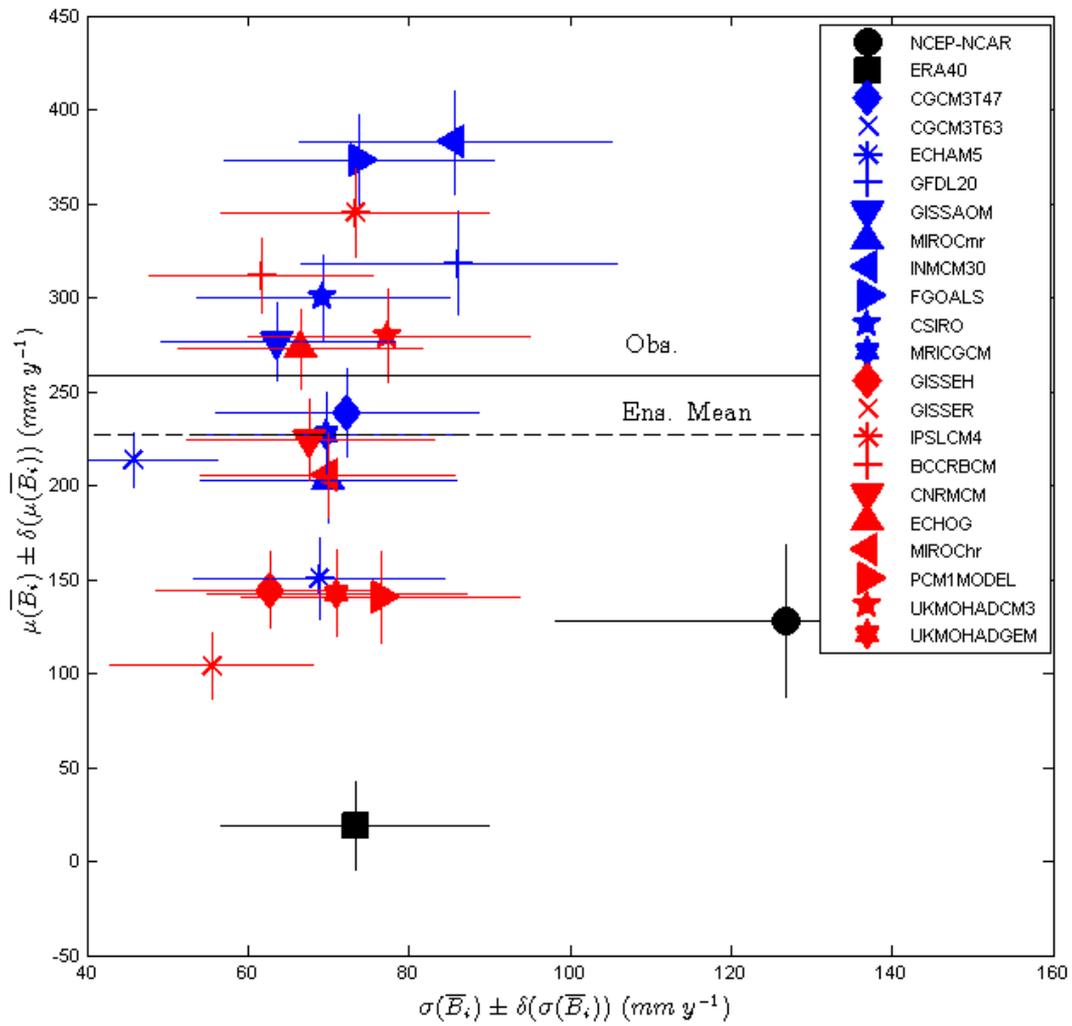

1
Figure 3. Estimates of the interannual variability vs. estimates of the basin-integrated yearly accumulated water
balance (symbols) and their 95% confidence intervals (horizontal and vertical lines). Obs. Stands for long-term
averaged discharge at sea (about 6600 m³s⁻¹). Note that 100 mm y⁻¹ correspond to about 2500 m³s⁻¹ of equivalent
mean river discharge. Further details in the text.

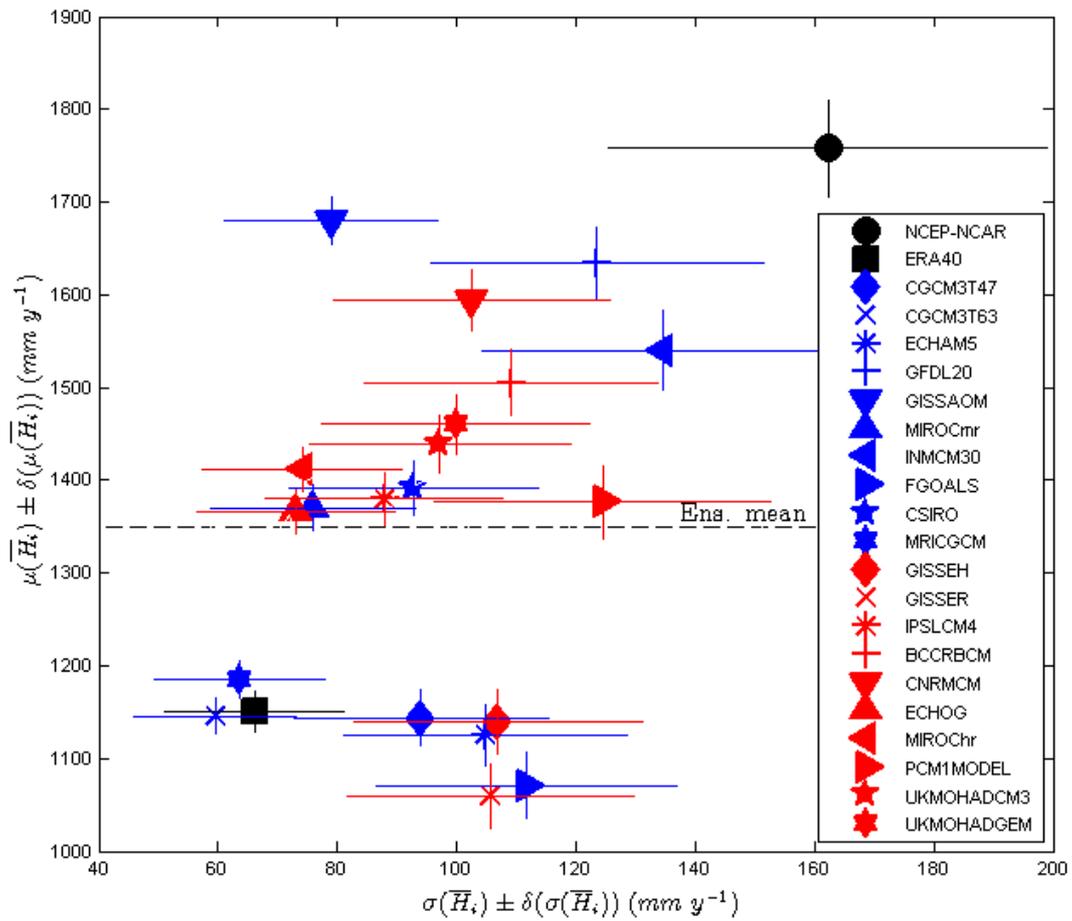

2 **Figure 4. Estimates of the interannual variability vs. estimates of the basin-integrated yearly accumulated**
3 **strength of the hydrological cycle (symbols) and their 95% confidence intervals (horizontal and vertical lines).**
4 **Note that 100 mm y$^{-1}$ correspond to about 2500 m$^3$s$^{-1}$of equivalent mean river discharge. Further details in the**
5 **text.**
6

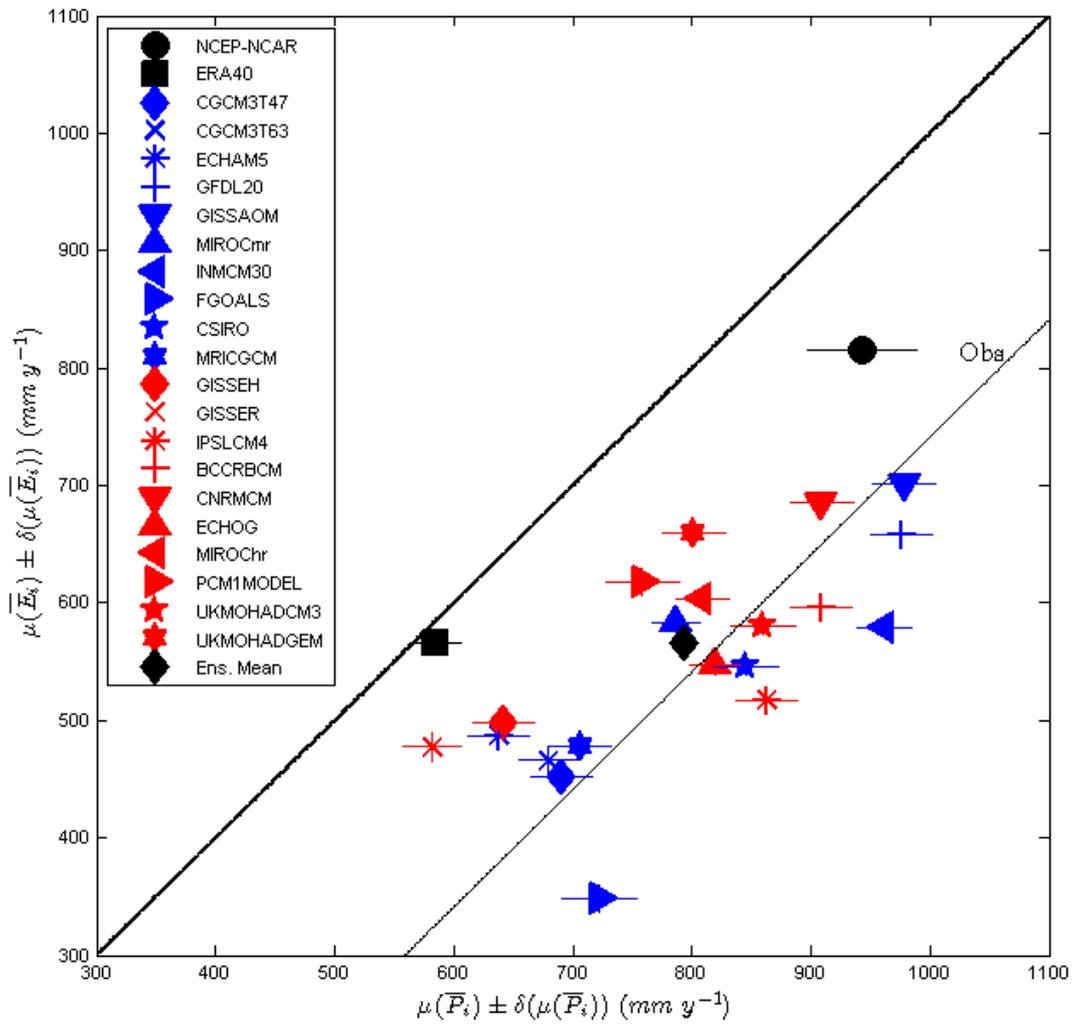

2  **Figure 5. Estimates of the basin integrated yearly accumulated precipitation vs. estimates of the basin-integrated**
3  **yearly accumulated evaporation (symbols) and their 95% confidence intervals (horizontal and vertical lines).**
4  **The bisectrix gives the zero water balance case. Obs. stands for long-term averaged discharge at sea (about 6600**
5  **$m^3s^{-1}$). Note that 100 $mm\ y^{-1}$ of precipitation correspond to about 2500 $m^3s^{-1}$ of equivalent mean river discharge.**
6  **Further details in the text.**

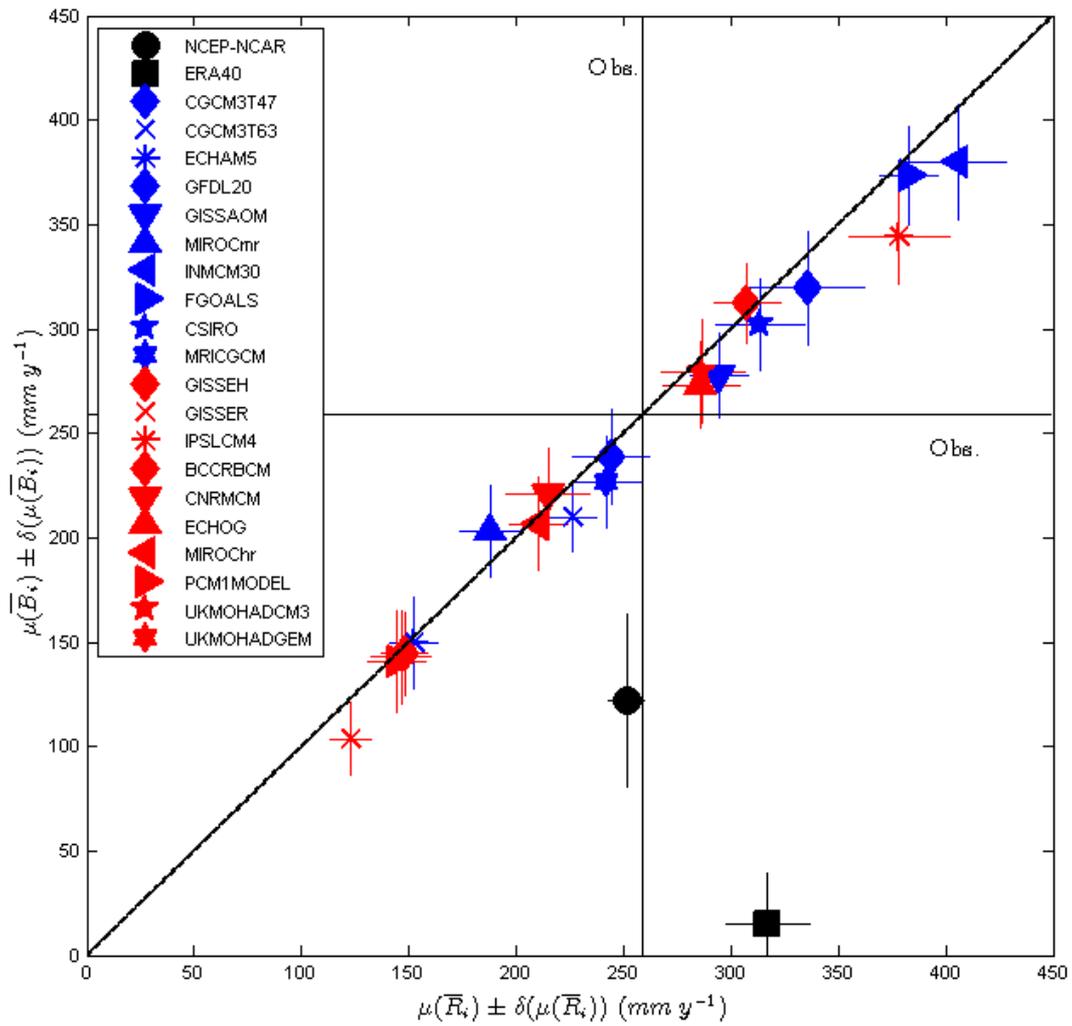

2  **Figure 6. Estimates of the basin integrated yearly accumulated runoff vs. estimates of the basin-integrated**
3  **yearly accumulated water balance (symbols) and their 95% confidence intervals (horizontal and vertical lines).**
4  **Obs. stands for long-term averaged discharge at sea (about 6600 $m^3s^{-1}$). The bisectrix indicates the theoretical**
5  **constraint all datasets should obey to. Note that 100 $mm\ y^{-1}$ of net water balance (or of runoff) correspond to**
6  **about 2500 $m^3s^{-1}$ of equivalent mean river discharge. Further details in the text.**

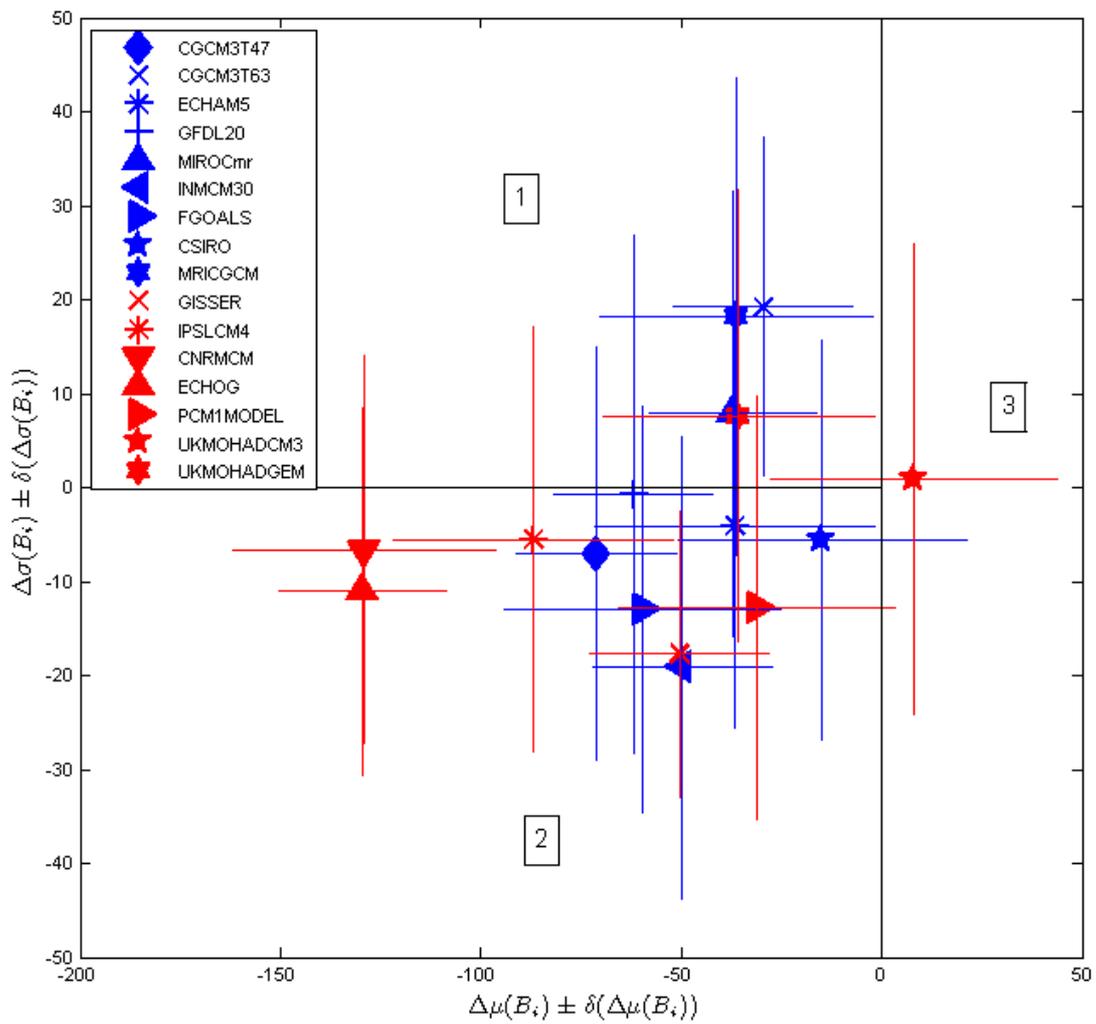

Figure 7. Change between the statistics of the 2161-2200 and the 1961-2000 basin-integrated annual accumulated
water balance. Estimates of the basin integrated yearly accumulated change in the water balance vs. change in its
interannual variability (symbols) and their 95% confidence intervals (horizontal and vertical lines). Note that
100 mm y$^{-1}$ of net water balance correspond to about 2500 m$^3$s$^{-1}$ of equivalent mean river discharge. Further
details in the text.

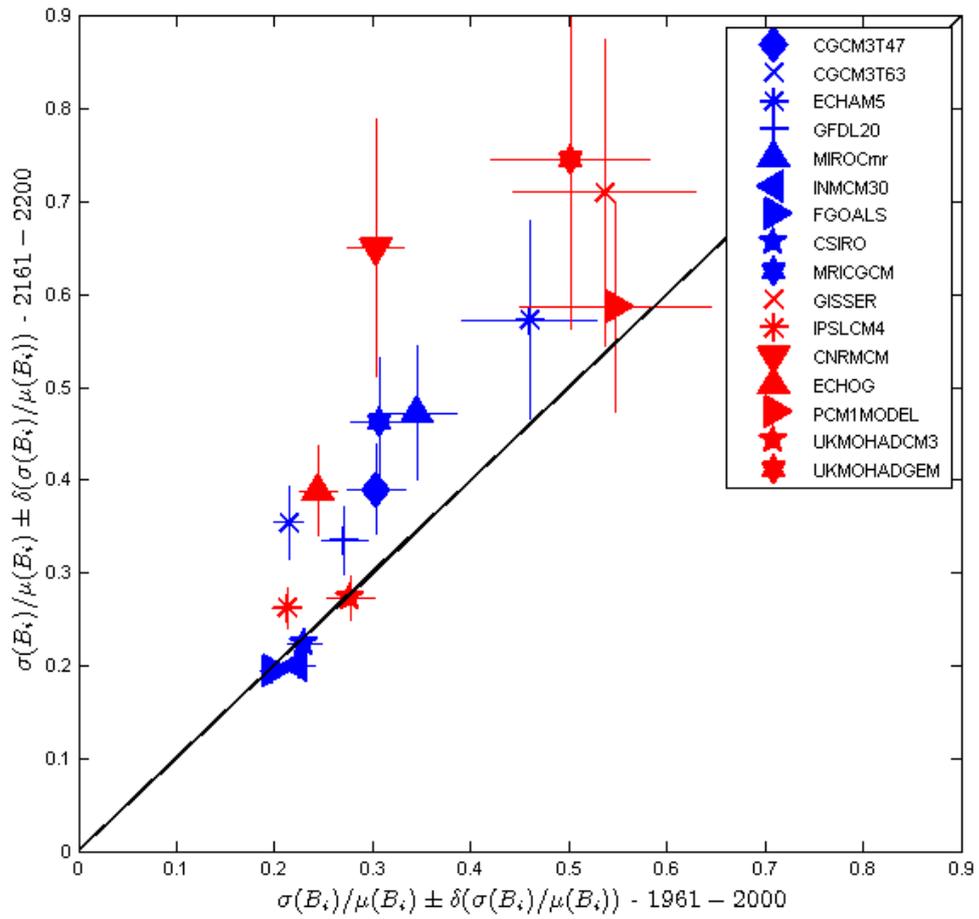

2 **Figure 8. Change between the statistics of the 2161-2200 and the 1961-2000 basin-integrated annual accumulated**
3 **water balance. Estimates of the 1961-2000 vs. the 2161-2200 ratio between the interannual variability and the**
4 **basin integrated yearly accumulated water balance (symbols) and their 95% confidence intervals (horizontal and**
5 **vertical lines). Further details in the text.**

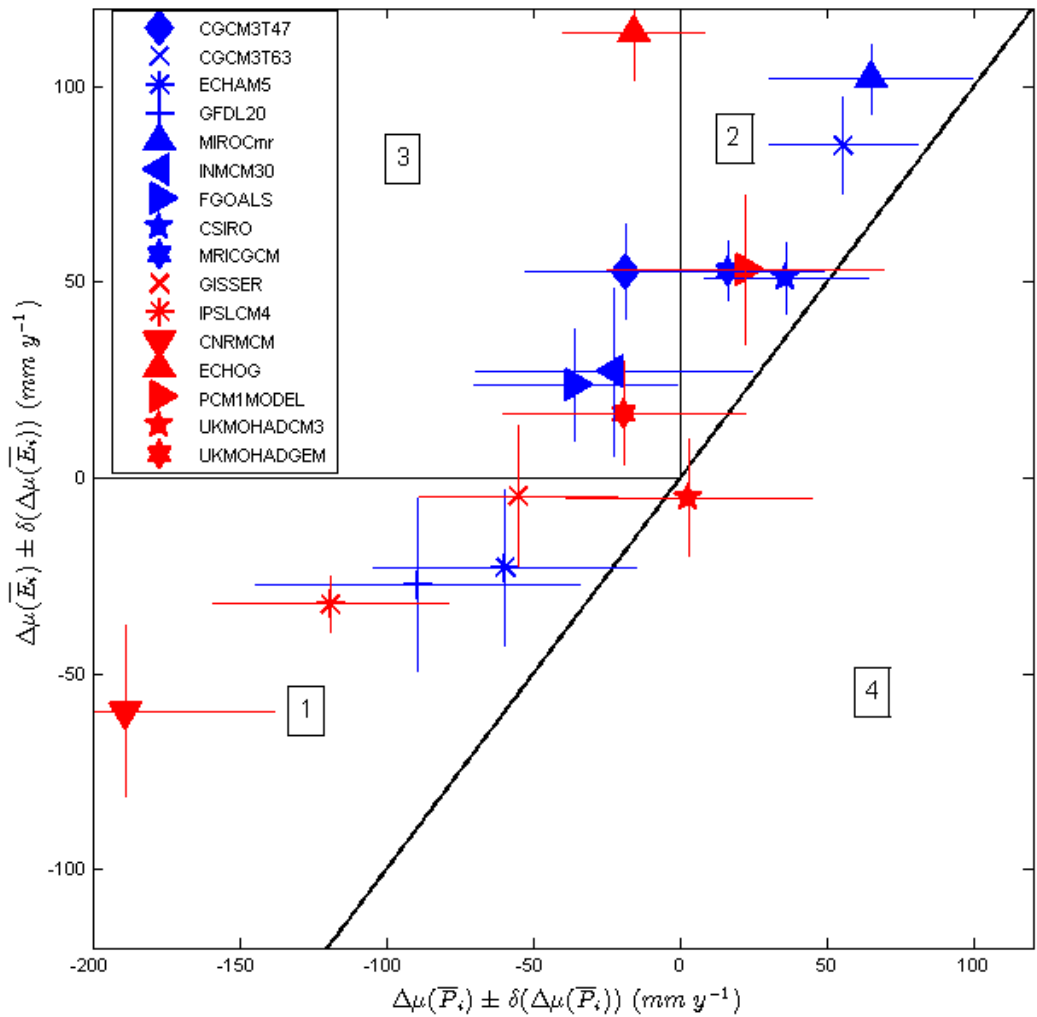

1 
2 **Figure 9. Change between the statistics of the 2161-2200 and the 1961-2000 basin-integrated annual accumulated**
3 **water balance. Estimates of the basin-integrated yearly accumulated change in precipitation vs. change in**
4 **evaporation (symbols) and their 95% confidence intervals (horizontal and vertical lines). Note that 100 mm y$^{-1}$ of**
5 **evaporation or precipitation correspond to about 2500 m$^3$s$^{-1}$ of equivalent mean river discharge. Further details**
6 **in the text.**

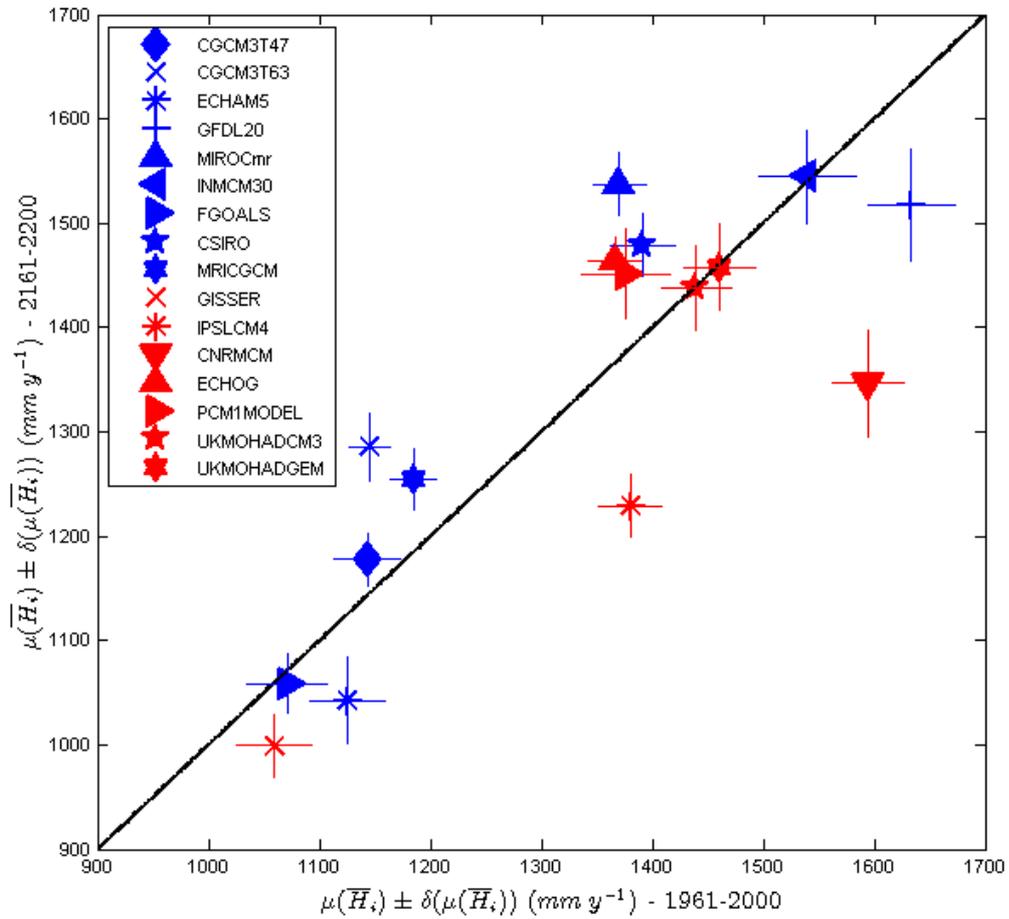

1
2 **Figure 10. Change between the statistics of the 2161-2200 and the 1961-2000 basin-integrated annual**
3 **accumulated total hydrological cycle. Estimates of the 1961-2000 vs. 2161-2200 basin-integrated yearly**
4 **accumulated strength of the hydrological cycle (symbols) and their 95% confidence intervals (horizontal and**
5 **vertical lines). Further details in the text.**

Table 1. An overview of the IPCC models. The horizontal resolution is expressed in terms of truncation (T) for spectral models (with T47 ≈ 2.8°, T63 ≈ 2.8°, T63 ≈ 2.8°); for Z coordinates, the letter before the number of levels indicates whether the vertical coordinate is height (z), pressure normalized with surface pressure (σ) or hybrid (h). For further information, refer to the PCMDI/CMIP3 web site http://www.pcmdi.llnl.gov.

| Model (Reference) | Institution | Trunc. (Lon x Lat) | Z |
|---|---|---|---|
| BCCRBCM Furevik et al. (2003) | Bjerknes Center for Climate Research, *Norway* | T63 | h31 |
| CNRMCM Salas-Mélia et al. (2005) | Mètèo France, *France* | T63 | h45 |
| CGCM3T47, CGCM3T63 Kim et al. (2002) | CCCma, *Canada* | T47(T63) | z31 |
| CSIRO Gordon et al. (2002) | CSIRO, *Australia* | T63 | h18 |
| ECHAM5 Jungclaus et al. (2006) | Max Planck Inst., *Germany* | T63 | h31 |
| ECHOG Min et al. (2005) | MIUB, METRI, and M&D, *Germany/Korea* | T30 | h19 |
| FGOALS Yu et al. (2004) | LASG, *China* | 2.8° x 2.8° | σ26 |
| GFDL20 Delworth et al. (2005) | GFDL, *USA* | 2.5° x 2.0° | h24 |
| GISSAOM Lucarini and Russell (2002) | NASA-GISS, *USA* | 4° x 3° | h12 |
| GISSEH, GISSER Schmidt et al. (2005) | NASA-GISS, *USA* | 4° x 5° | σ20 |
| INMCM30 Volodin and Diansky (2004) | Inst. of Num. Math., *Russia* | 5° x 4° | σ21 |
| IPSLCM4 Marti et al. (2005) | IPSL, *France* | 2.4° x 3.75° | h19 |
| MIROChr, MIROCmr K-1 mod. Dev. (2004) | CCSR/NIES/FRCGC, *Japan* | T106(*hires*) T42(*medres*) | σ56(*hires*) σ20(*medres*) |
| MRICGC Yukimoto and Noda (2002) | Meteorological Research Institute,*Japan* | T42 | h30 |
| PCM1MODEL Meehl et al. (2004) | National Center for Atmospheric Research, *USA* | T42 | h26 |
| UKMOHADCM3 Johns et al. (2003) | Hadley Centre for Climate Prediction and Research / Met Office, *UK* | 2.75° x 3.75° | h19 |
| UKMOHADGEM3 Johns et al. (2006) | Hadley Centre for Climate Prediction and Research / Met Office, *UK* | 1.25° x 1.875° | h38 |

1  **Table 2.** Linear time correlations between the yearly accumulated basin-integrated evaporation (E), precipitation (P),
2  and water balance (B) time series in the 1961-2000 and 2161-2200 time frames. Statistically significant correlations are
3  indicated in bold.
4

| Datasets | C(E,P) 1961-2000 | C(E,P) 2161-2200 |
|---|---|---|
| NCEP-NCAR | **0,75** | |
| ERA40 | -0,22 | |
| CGCM3T47 | **0,39** | 0,31 |
| CGCM3T63 | 0,30 | **0,66** |
| ECHAM5 | **0,59** | **0,75** |
| GFDL20 | **0,59** | **0,80** |
| GISSAOM | **0,45** | |
| MIROCmr | 0,15 | **0,42** |
| INMCM30 | **0,58** | **0,80** |
| FGOALS | **0,60** | **0,52** |
| CSIRO | **0,57** | **0,72** |
| MRICGCM | -0,19 | 0,10 |
| GISSEH | **0,69** | |
| GISSER | **0,72** | **0,81** |
| IPSLCM4 | **0,56** | **0,77** |
| BCCRBCM | **0,73** | |
| CNRMCM | **0,68** | **0,88** |
| ECHOG | 0,13 | **0,35** |
| MIROChr | 0,15 | |
| PCM1MODEL | **0,69** | **0,82** |
| UKMOHADCM3 | **0,45** | **0,65** |
| UKMOHADGEM | **0,49** | **0,74** |

# References

Accadia, C., S. Mariani, M. Casaioli, A. Lavagnini, and A. Speranza (2003) Sensitivity of precipitation forecast skill scores to bilinear interpolation and a simple nearest-neighbour average method on high-resolution verification grids, Wea. Forecast., 18, 918–932.

Alpert, P., M. Tsidulko, S. Krichak, and U. Stein (1996) A multi-stage evolution of an ALPEX cyclone, Tellus A,**48**, 209–222

Amenu, G., and P. Kumar, (2005) NVAP and Reanalysis-2 Global Precipitable Water Products: Intercomparison and Variability Studies, Bull. Am. Meteor. Soc. **86**, 245-256.

Anderson, C. J., R. W. Arritt, E.S. Takle, Z. Pan, W. J. Gutowski, Jr., F. O. Otieno, R. da Silva, D. Caya, J. H. Christensen, D. Luthi, M. A. Gaertner, C. Gallardo, F. Giorgi, S.-Y. Hong, C. Jones, H.-M. H. Juang, J. J. Katzfey, W. M. Lapenta, R. Laprise, J. W. Larson, G. E. Liston, J. L. McGregor, R. A. Pielke, Sr., J. O. Roads, and J. A. Taylor (2003) Hydrological processes in regional climate model simulations of the central United States flood of June-July 1993, J. Hydrometeor. 4, 584-598

Artale V, S. Calmanti, and A. Sutera (2002) Thermohaline circulation sensitivity to intermediate-level anomalies, Tellus A 54(2): 159-174

Becker A., and U. Grunewald (2003) Disaster Management: Flood Risk in Central Europe, Science 300, 1099

Betts, A. K., P. Viterbo, and E. Wood (1998) Surface energy and water balance for the Arkansas-Red River basin from the ECMWF reanalysis. J. Clim., 11, 2881-2897

Betts, A. K., J. H. Ball, and P. Viterbo (1999) Basin-scale surface water and energy budgets for the Mississippi from the ECMWF reanalysis. J. Geophys. Res., 104(D16), 19.293-19.306

Betts, A. K., J. H. Ball, and P. Viterbo (2003) Water and energy budgets for the Mackenzie river basins from ERA-40. J. Hydrometeor., 4, 1194-1211

Calmanti S, V. Artale, and A. Sutera (2006), North Atlantic MOC variability and the Mediterranean Outflow: a box-model study, Tellus A 58 (3), 416–423.

Deidda, R. (1999) Multifractal analysis and simulation of rainfall fields in space, Physics and Chemistry of the Earth (B), 24, 73-78

Deidda, R. (2000) Rainfall downscaling in a space-time multifractal framework, Water Resources Research, 36, 1779-1784.

Delworth, T. L., A. J. Broccoli, A. Rosati, R. J. Stouffer, V. Balaji, J. A. Beesley, W. F. Cooke,K. W. Dixon, J. Dunne, K. A. Dunne,J. W. Durachta, K. L., Findell, P. Ginoux, A. Gnanadesikan, C. T. Gordon, C. T., S. M. Gri_es, R. Gudgel, M. J. Harrison, I. M. Held, R, S. Hemler, L. W. Horowitz, S. A. Klein, T. R. Knutson, P. J. Kushner, A. R.

Langenhorst, H. C. Lee, S. J. Lin, J. Lu, S. L. Malyshev, P. C. D. Milly, V. Ramaswamy, J. Russell, M. D. Schwarzkopf, E. Shevliakova, J. J. Sirutis, M. J. Spelman, W. F. Stern, M. Winton, A. T.Wittenberg, B. Wyman, F. Zeng, and R. Zhang (2005) GFDL's CM2 global coupled climate models – Part 1: formulation and simulation characteristics, J. Climate 19: 643-674.

Elsner, J. B., V. K. Gupta, S. Lovejoy, V. Lucarini, A. B. Murray, A. S. Sharma, S. Tebbens, A. A. Tsonis, and D. Vassiliadis (2007) Twenty Years of Nonlinear Dynamics in Geosciences, Eos Trans. AGU, 88(3), 29, doi:10.1029/2007EO030006

Furevik T., M. Bentsen, H. Drange, I. K. T. Kindem, N. G. Kvamsto, and A. Sorteberg (2003) Description and evaluation of the Bergen climate model: ARPEGE coupled with MICOM, Climate Dynamics 21, 27-51.

Giorgi, F., and X. Bi (2005a), Regional changes in surface climate interannual variability for the 21st century from ensembles of global model simulations, Geophys. Res. Lett., 32, L13701, doi:10.1029/2005GL023002.

Giorgi, F., and X. Bi (2005b), Updated regional precipitation and temperature changes for the 21st century from ensembles of recent AOGCM simulations, Geophys. Res. Lett., 32, L21715, doi:10.1029/2005GL024288

Giorgi, F. (2006) Climate change hot-spots. Geophys. Res. Lett. 33: L08707

Gordon, H. B., L. D. Rotstayn, J. L. McGregor, , M. R. Dix, E. A. Kowalczyk, S. P. O'Farrell, L. J. Waterman, A. C. Hirst, S. G. Wilson, M. A. Collier, I. G. Watterson, and T. I. Elliott (2002) The CSIRO Mk3 Climate System Model [Electronic publication]. Aspendale: CSIRO Atmospheric Research. (CSIRO Atmospheric Research technical paper; no. 60). 130 pp. ( http://www.dar.csiro.au/publications/gordon_2002a.pdf )

Gutowski W. J. Jr., Y. Chen, and Z. Ötles (1997) Atmospheric Water Vapor Transport in NCEP–NCAR Reanalyses: Comparison with River Discharge in the Central United States, Bull. Amer. Meteor. Soc. 78, 1957–1969

Hagemann, S., B. Machenhauer, R. Jones, O. B. Christensen, M. Déqué, D. Jacob, P. L. Vidale (2004) Evaluation of water and energy budgets in regional climate models applied over Europe, Climate Dynamics, 23, 547-567

Hagemann, S., K. Arpe, and L. Bengtsson (2005) Validation of the hydrological cycle of ERA-40. ERA-40 Project Report Series No. 24, ECMWF, England, 46 pp.

Held, I. M., and B. J. Soden (2006) Robust responses of the hydrological cycle to global warming, Journal of Climate 19, 5686-5699

Hirschi, M., S. I. Seneviratne, and C. H. Schär (2006) Seasonal Variations in Terrestrial Water Storage for Major Mid-latitude River Basins. J. Hydrometeor., 7, 39-60

IPCC (2007) Climate Change 2007: The Physical Science Basis. Contribution of Working Group I to the Fourth Assessment Report of the Intergovernmental Panel on Climate Change [Solomon, S., D. Qin, M. Manning, Z. Chen, M. Marquis, K.B. Averyt, M. Tignor and H.L. Miller (eds.)]. Cambridge University Press, Cambridge, United Kingdom and New York, NY, USA, 996 pp

Johns T.C., J. M. Gregory, W. J. Ingram, C. E. Johnson, A. Jones, J. A. Lowe, J. F. B. Mitchell, D. L. Roberts, B. M. H. Sexton, D. S. Stevenson, S. F. B. Tett, and M. J. Woodage (2003) Anthropogenic climate change for 1860 to 2100 simulated with the HadCM3 model under updated emissions scenarios, Clim. Dyn. 20, 583-612

Johns T.C., C. F. Durman, H. T. Banks, M. J. Roberts, A. J. McLaren, J. K. Ridley, C. A. Senior, K. D. Williams, A. Jones, G. J. Rickard, S. Cusack, W. J. Ingram, M. Crucifix, D. M. H. Sexton, M. M. Joshi, B-W. Dong, H. Spencer, R. S. R. Hill, J. M. Gregory, A.B. Keen, A. K. Pardaens, J. A. Lowe, A. Bodas-Salcedo, S. Stark, and Y. Searl (2006) The new Hadley Centre climate model HadGEM1: Evaluation of coupled simulations, J. Climate 19, 1327-1353

Jungclaus, J., M. Botzet, H. Haak, N. Keenlyside, J.-J. Luo, M. Latif, J. Marotzke, J. Mikolajewicz,, and E. Roeckner (2006) Ocean circulation and tropical variability in the AOGCM ECHAM5/MPI-OM, J. Climate, 19, 3952-3972

K-1 model developers (2004) K-1 coupled model (MIROC) description, K-1 technical report, 1, H. Hasumi and S. Emori (eds.), Center for Climate System Research, University of Tokyo, 34pp.

Kim, S.-J., G. M. Flato, G. J. Boer, and N. A. McFarlane (2002) A coupled climate model simulation of the Last Glacial Maximum, Part 1: transient multi-decadal response. Climate Dynamics, 19, 515-537.

Kistler, R., E. Kalnay, W. Collins, S. Saha, G. White, J. Woollen, M. Chelliah, W. Ebisuzaki, M. Kanamitsu, V. Kousky, H. van den Dool, R. Jenne, M. Fiorino (2001) The NCEP-NCAR 50-Year Reanalysis: Monthly Means CD-ROM and Documentation. Bull. Amer. Meteor. Soc., 82, 247-268

Kuhn T. S. (1970) The Structure of Scientific Revolutions, University of Chicago Press, Chicago

Lau, K. M., J. H. Him , Y. Sud, (1996) Intercomparison of Hydrologic Processes in AMIP GCMs. Bull. Amer. Meteor. Soc., 77, 2209-2227

Lovejoy, S., and D. Schertzer (2006) Multifractals, cloud radiances and rain, 2006: J. of Hydrol.. **322,** 59-88.

Lucarini, V. (2002) Towards a definition of climate science. Int. J. Environment and Pollution 18: 409-414.

Lucarini V. and G. L. Russell (2002) Comparison of mean climate trends in the northern hemisphere between National Centers for Environmental Prediction and two atmosphere-ocean model forced runs. J. Geophys. Res., 107 (D15), 10.1029/2001JD001247

Lucarini V., T. Nanni, A. Speranza (2006) Statistical Properties of the seasonal cycle in the Mediterranean area, Il Nuovo Cimento C 29, 21-31

Lucarini, V., S. Calmanti, A. Dell'Aquila, P.M. Ruti and A. Speranza (2007a) Intercomparison of the northern hemisphere winter mid-latitude atmospheric variability of the IPCC models, Climate Dynamics, **28**, 829-848, doi: 10.1007/s00382-006-0213-x

Lucarini V., R. Danihlik, I. Kriegerova, and A. Speranza, (2007b) Does the Danube exist? Versions of reality given by various regional climate models and climatological datasets, Journal of Geophysical Research, 112, D13103, doi: 10.1029/2006JD008360

Meehl, G.A., W. M. Washington, C. Ammann, J. M. Arblaster, T. M. L. Wigley, and C. Tebaldi (2004) Combinations of natural and anthropogenic forcings and 20th century climate. J. Climate, 17, 3721—3727

Marti, O. et al. (2005) The new IPSL climate system model: IPSL-CM4, Tech. rep., Institut Pierre Simon Laplace des Sciences de l'Environnement Global, IPSL, Case 101, 4 place Jussieu, Paris, France.

Min, S.-K., S. Legutke, A. Hense, and W.-T. Kwon (2005) Internal variability in a 1000-year control simulation with the coupled climate model ECHO-G. Part I. Near-surface temperature, precipitation and mean sea level pressure. Tellus, 57A, 605-621.

Okabe A., B. Boots, K. Sugihara, and S. Nok Chiu (2000) Spatial Tessellations - Concepts and Applications of Voronoi Diagrams, John Wiley and Sons, New York, USA

Peixoto, J.P. and A. H. Oort (1992) Physics of Climate, American Institute of Physics, New York,

Rahmstorf, S. (1998) Influence of Mediterranean Outflow on climate. Eos **79**, 281–282.

Roads J. O., S.-C. Chen, A. K. Guetter, and K. P. Georgakakos (1994) Large-Scale Aspects of the United States Hydrologic Cycle**,** Bull. Amer. Meteor. Soc. 75, 1589–1610

Roads, J. and A. Betts (1999) NCEP-NCAR and ECMWF Reanalysis Surface Water and Energy Budgets for the Mississippi River Basin. J. Hydrometeor, 1, 88-94

Salas-Mélia, D., F. Chauvin, M. Déqué, H. Douville, J.F. Gueremy, P. Marquet, S. Planton, J.F. Royer, and S. Tyteca, (2005) Description and validation of the CNRM-CM3 global coupled model, submitted to Climate Dynamics.

Schmidt, G. A., R. Ruedy, J. E. Hansen, I. Aleinov, N. Bell, M. Bauer, S. Bauer, B. Cairns, V. Canuto, Y. Cheng, A. DelGenio, G. Faluvegi, A. D. Friend, T. M. Hall, Y. Hu, M. Kelley, N. Y. Kiang, D. Koch, A. A. Lacis, J. Lerner, K. K. Lo, R. L. Miller, L. Nazarenko, V. Oinas, J. Perlwitz, J. Perlwitz, D. Rind, A. Romanou, G. L. Russell, M. Sato, D. T. Shindell, P. H. Stone, S. Sun, N. Tausnev, D. Thresher, and M.-S. Yao (2005) Present day atmospheric simulations using GISS Model E: Comparison to in-situ, satellite and reanalysis data. J. Climate 19: 153-192.

Simmons A. J. and J. K. Gibson (2000) The ERA-40 Project Plan, ERA-40 Project Report Series No. 1, European Centre for Medium-Range Weather Forecasts, Reading, United Kingdom

Speranza A. (2002) The hydrological cycle of the Mediterranean Basin, Proceedings of the Interreg IIC Conference *Drought-Monitoring, Mitigation, Effects*, Villasimius (Cagliari, Italy), 21-23 September 2000, Eds G. Monacelli and E. Giusta., pp. 103-106

Speranza A., A. Buzzi, A. Trevisan, and P. Malguzzi (1985) A theory of deep cyclogenesis in the lee of the Alps: modifications of baroclinic instability by localized topography, J. Atmos. Sci. 42, 1521-1535.

Stohl A. and P. James (2004) A Lagrangian Analysis of the Atmospheric Branch of the Global Water Cycle. Part I: Method Description, Validation, and Demonstration for the August 2002 Flooding in Central Europe, Journal of Hydrometeorology, 5, 656-678

Tessier, Y., S. Lovejoy, and D. Schertzer (1993) Universal Multifractals: Theory and Observations for Rain and Clouds, J. Appl. Meteor. 32, 223-250

Tibaldi S., A. Buzzi and A. Speranza (1990) Orographic cyclogenesis. In: C.W. Newton, E.O. Holopainen (eds.) Extratropical Cyclones-The Eric Palmen Memorial Volume. Boston: American Meteorological Society, 107–127

Troccoli A. and P. Kallberg (2004), Precipitation correction in the ERA-40 reanalysis, ERA-40 Project Report Series, 13

Tsonis A. A. and J. B. Elsner (Eds.), 2007: Nonlinear Dynamics in Geosciences, Springer, New York

Volodin E.M. and N.A. Diansky (2004) El-Nino reproduction in coupled general circulation model of atmosphere and ocean. Russian meteorology and hydrology 12, 5-14

Wilks, D.S. (1997) Resampling hypothesis tests for autocorrelated field. J.Climate 10, 65-82.

Yu Y., X. Zhang, Y. Guo (2004) Global coupled ocean- atmosphere general circulation models in LASG/IAP. Adv. Atmos. Sci., 21, 444-455.

Yukimoto, S. and A. Noda (2002) Improvements of the Meteorological Research Institute Global Ocean-atmosphere Coupled GCM (MRI-CGCM2) and its climate sensitivity, Tech. Rep. 10, NIES, Japan.